\documentclass[pra,reprint,superscriptaddress,floatfix,aps]{revtex4-2}
\usepackage{amsmath}
\usepackage{amssymb}
\usepackage{amsfonts}
\usepackage{graphicx}
\usepackage{physics}
\usepackage{array}
\usepackage{multirow}
\usepackage{tikz}
\usepackage{enumitem}
\usepackage[colorlinks,linktocpage,bookmarks=false,citecolor=blue,linkcolor=red,urlcolor=blue]{hyperref}

\def\be{\begin{equation}} \def\ee{\end{equation}}
\def\bea{\begin{eqnarray}} \def\eea{\end{eqnarray}}

\begin{document}

\title{Directional quantum walks of two bosons on the Hatano-Nelson lattice}

\author{Sk Anisur}
\affiliation{Harish-Chandra Research Institute, a CI of Homi Bhabha National Institute, Chhatnag Road, Jhunsi, Allahabad 211019}
\affiliation{Homi Bhabha National Institute, Training School Complex, Anushakti Nagar, Mumbai 400 094, India}
\date{\today}

\author{Kartik Singh}
\affiliation{Indian Institute of Science Education and Research, Pune 411008, India}

\author{Sayan Choudhury}
\email{sayanchoudhury@hri.res.in}
\affiliation{Harish-Chandra Research Institute, a CI of Homi Bhabha National Institute, Chhatnag Road, Jhunsi, Allahabad 211019}
\affiliation{Homi Bhabha National Institute, Training School Complex, Anushakti Nagar, Mumbai 400 094, India}
\date{\today}

\date{\today}
\begin{abstract}
We theoretically investigate the interplay of interactions and non-Hermiticity in the dynamics of two bosons on the one-dimensional Hatano-Nelson lattice with non-reciprocal tunneling. We find that the non-reciprocity in the tunneling leads to the formation of an asymmetric density cone during the time-evolution of the system; the degree of asymmetry can be controlled by tuning the non-reciprocity parameter, $\delta$. Next, we analyze the dynamics of this system in the presence of a static external force, $F$, and demonstrate that non-Hermiticity leads to asymmetric two-particle Bloch oscillations. Interestingly, when $F=0$ ($F \ne 0$), strong interactions lead to the formation of an inner density-cone (density-hourglass) structure; this inner structure also becomes asymmetric in the presence of non-Hermiticity. We further analyze the spatial correlations that develop between the two bosons, and establish that the system exhibits non-reciprocal bunching (anti-bunching)  in the presence of weak (strong) interactions. Finally, we examine the growth of the quantum Fisher information, $F_Q$, with time, and demonstrate that $F_Q$ decreases with increasing $\delta$. Interestingly however, for both one- and two-particle walks, the rapid growth of $F_Q$ characteristic of the Hermitian case, with an effective power law $F_Q \propto t^{\alpha}$, $\alpha \simeq 3$, persists in the presence of non-reciprocal tunneling.
\end{abstract}
\maketitle
\section{Introduction} 
\label{sec:Intro}

The quantum walk (QW) of multiple bosons provides a powerful setting for investigating the interplay of interactions and statistics~\cite{aharonov1993quantum,kempe2003quantum,venegas2012quantum,mulken2011continuous}. In particular, continuous-time multi-particle bosonic QWs have been employed to study the emergence of quantum correlations due to Hanbury Brown–Twiss interference~\cite{lahini2012quantum,preiss2015strongly}, and develop protocols for universal computation~\cite{farhi1998quantum,childs2009universal,childs2013universal,qiang2024quantum} and quantum-enhanced sensing of weak forces~\cite{cai2021multiparticle,yang2024controllable}. Furthermore, QWs provide valuable insights into the dynamics of entanglement~\cite{beggi2018probing} and quantum magic resources~\cite{moca2025non}. QWs have now been realized in a variety of platforms - ranging from ultracold atoms~\cite{tai2017microscopy,xie2020topological,young2022tweezer,chen2024quantum,weckesser2025realization} to photonic systems~\cite{peruzzo2010quantum,kitagawa2012observation,zhou2024multi} and superconducting qubit processors~\cite{yan2019strongly,gong2021quantum}.\\

Most studies on QWs so far have focused on the time evolution of a quantum system governed by a Hermitian Hamiltonian. However, in recent years, the properties of non-Hermitian Hamiltonians have emerged as a topic of intense interest~\cite{gong2018topological,kawabata2019symmetry,ashida2020non,okuma2023non,banerjee2023non,fruchart2021non,el2018non}. These Hamiltonians naturally emerge in the context of open classical and quantum systems, and they exhibit a wide array of remarkable phenomena that don't have any Hermitian counterparts. Spectacular examples of intrinsically non-Hermitian phenomena include the existence of exceptional points~\cite{bergholtz2021exceptional,ding2022non,li2023exceptional}, where both eigenvectors and eigenvalues coalesce, and the non-Hermitian skin effect (NHSE), characterized by a macroscopic accumulation of eigenstates towards the system boundary~\cite{yao2018edge,zhang2022review}. The NHSE has now been extensively studied theoretically and experimentally in a variety of platforms~\cite{kunst2018biorthogonal,lee2019anatomy,helbig2020generalized,martinez2018non,borgnia2020non,zhang2020correspondence,yi2020non,okuma2020topological,kawabata2023entanglement}.\\

Given the above background, it is natural to investigate QWs in a non-Hermitian setting. In this context, we note that several theoretical and experimental works have demonstrated that single-particle QWs provide a powerful tool to probe the NHSE in Hatano-Nelson lattices characterized by non-reciprocal tunneling~\cite{hatano1996localization,hatano1997vortex,hatano1998non}. Furthermore, these works have been extended to explore Bloch oscillations in the presence of a static electric field~\cite{graefe2016quasiclassical,longhi2016bloch,peng2022manipulating} and localization due to random~\cite{claes2021skin,zhang2023bulk,kokkinakis2024anderson,shang2025spreading,wang2025unified} and quasi-periodic disorder~\cite{padhi2024quasi,wang2025coexistence,Chakrabarty2023skin,Chakrabarty2025fate}. However, two-particle QWs in non-Hermitian systems have not received much attention. 

In this work, we address this gap by investigating the QW of two interacting bosons on a Hatano-Nelson lattice, both in the absence and in the presence of a static force. We find that the interplay of non-Hermiticity, strong interactions, and bosonic statistics can lead to intriguing new features both in the spread of the density and the spatial correlations. Finally, we demonstrate that as in the Hermitian case, the quantum Fisher information, $F_Q$ grows as $F_Q \propto t^{\alpha}$ with $\alpha \simeq 3$ for both single-particle and two-particle non-Hermitian QWs. Our results pave the path towards understanding the rich physics of multi-particle QWs in non-Hermitian systems.

This work is organized as follows. We describe the model and summarize known results about this system in Sec.~\ref{sec:Model}. We explore the directional QWs of two bosons in the absence of an external field in Sec.~\ref{sec:NHSE}. We then examine the fate of Bloch oscillations in the presence of non-Hermiticity in Sec.~\ref{sec:Bloch}. Finally, we examine the growth of the QFI for these non-Hermitian QWs in Sec.~\ref{sec:QFI}, and conclude with a summary of our results and an outlook for future work in Sec.~\ref{sec:Conclusion}. 

\section{Model and Summary of Previous works}
\label{sec:Model}
We study the one-dimensional Hatano-Nelson-Bose-Hubbard (HNBH) model featuring non-reciprocal tunneling in the presence of a dc field~\cite{zheng2024exact,ibarra2025autoregressive}:

\begin{align}
\hat{H} = 
& \sum_{j=0}^{L-1} -J\left[ (1 - \delta)\, \hat{a}_{j+1}^{\dagger} \hat{a}_j + (1 + \delta)\, \hat{a}_j^{\dagger} \hat{a}_{j+1} \right] \nonumber \\
& + \frac{U}{2} \sum_{j=0}^{L} \, \hat{a}_j^{\dagger} \hat{a}_j^{\dagger} \hat{a}_j \hat{a}_j + F \sum_{j=0}^{L} j \, \hat{a}_j^{\dagger} \hat{a}_j,
\end{align}
where $\hat{a}_j$ is the bosonic annihilation operator at site $j$, $J$ denotes the nearest-neighbor tunneling matrix element, $\delta$ represents the non-reciprocity parameter, $U$ is the on-site interaction strength, and $F$ quantifies the dc field strength. We set $J=1$ for the rest of this work, which sets the natural units of energy and time. Furthermore, we consider two kinds of archetypal initial states in this work: (a) when the two bosons are on neighboring sites at the center of the lattice: 
\begin{equation}
    \vert \psi_{\rm ini}^{(1)} \rangle = \hat{a}_{L/2}^{\dagger} \hat{a}_{L/2+1}^{\dagger} \vert 0 \rangle,
    \label{eq:IniState1}
\end{equation} 
and (b) when both bosons are on the same central site:
\begin{equation}
\vert \psi_{\rm ini}^{(2)} \rangle = \frac{1}{\sqrt{2}} \hat{a}_{L/2}^{\dagger} \hat{a}_{L/2}^{\dagger} \vert 0 \rangle. 
\label{eq:IniState2}
\end{equation}
Furthermore, since the non-Hermitian evolution does not conserve the norm, all observables computed in this work, including the QFI, are obtained from the normalized state:
\begin{equation}
\lvert \psi(t) \rangle = e^{-i\hat{H}t}\lvert
\psi(0) \rangle / \lVert e^{-i\hat{H}t}\lvert \psi(0) \rangle \rVert
\end{equation}
at each time $t$. We now proceed to review the known results on QWs in this model.\\

\begin{figure*}[t]
    \centering
    \includegraphics[width=\linewidth]{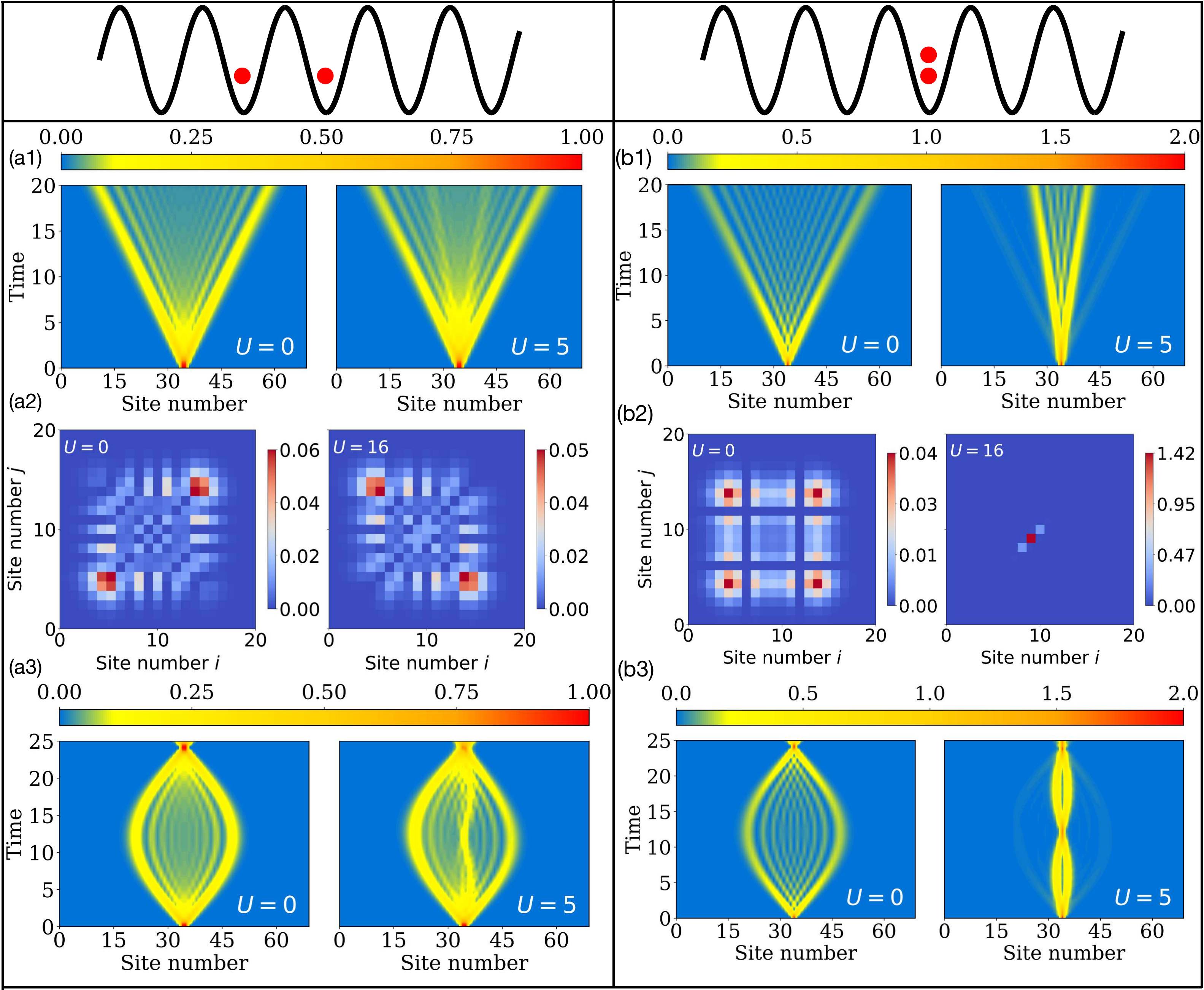}
    \caption{\textbf{Two-particle quantum walks in the Bose-Hubbard model:} The left and right panels show QWs for the $\vert \psi_{\rm ini}^{(1)}\rangle$ and $\vert \psi_{\rm ini}^{(2)}\rangle$ states, respectively. (a1)-(b1): The time evolution of the density in the absence of a dc field. A density cone is formed in the non-interacting limit ($U=0$), which fragments with increasing interaction strength. (a2)-(b2): The two-particle correlator, $\Gamma_{i,j}$ (Eq.~\ref{eq:Correlator}) computed at $t=\pi$. Increasing $U$ leads to a transition from bosonic bunching correlations to fermionic anti-bunching correlations for the $\vert \psi_{\rm ini}^{(1)}\rangle$ state. On the other hand, for the $\vert \psi_{\rm ini}^{(2)}\rangle$ state, greater doublon localization is induced by increasing $U$. (a3)-(b3): The time evolution of the density in the presence of a dc field ($F \ne 0$). Bloch oscillations are observed for both initial states when $U = 0$. Interactions induce an inner hourglass structure in the density profile. The hourglass structure is more prominent for the $\vert \psi_{\rm ini}^{(2)}\rangle$ state  due to doublon formation.}
    \label{Fig:1}
\end{figure*}

\newlist{tbullet}{itemize}{1}
\setlist[tbullet]{leftmargin=1em, label=\tiny\textbullet, itemsep=1pt, topsep=2pt, parsep=1pt}

\paragraph{Hermitian QWs in the absence of a dc field (F=0):} We begin by reviewing the Hermitian regime ($\delta=0$), where the model reduces to the well-known Bose-Hubbard (BH) model. QWs have been studied extensively both theoretically~\cite{khomeriki2010interaction,dias2007frequency,wiater2017two}  and experimentally~\cite{lahini2012quantum,preiss2015strongly} in the BH model. Most of these studies have focused on the bosonic lattice density distribution:
\begin{equation}
    \hat{n}_i (t) = \langle \psi (t) \vert \hat{a}_i^{\dagger} \hat{a}_i \vert \psi (t) \rangle.
    \label{eq:Density}
\end{equation}
In the absence of a dc field ($F=0$), the wavefunction of an initially localized particle spreads ballistically: $\hat{n}_i (t) =\left| J_i(2t) \right|^2$, where $J_i$ denote Bessel functions of the first kind~\cite{hartmann2004dynamics}. In the case of two interacting bosons ($U>0$), the time-evolution of the $\vert \psi_{\rm ini}^{(1)} \rangle$ and $\vert \psi_{\rm ini}^{(2)} \rangle$ states shows distinct behaviors. It is instructive to decompose the density into two parts $\hat{n}_i (t) = \hat{n}_i^{(1)} (t) + \hat{n}_i^{(2)} (t)$, where $\hat{n}_i^{(2)} (t) = \langle \psi(t) \vert \hat{a}_i^{\dagger} \hat{a}_i^{\dagger} \hat{a}_i \hat{a}_i \vert \psi (t) \rangle$ and  $\hat{n}_i^{(1)} (t) = \hat{n}_i (t) - \hat{n}_i^{(2)} (t)$. During the time evolution, the density profile fragments into two parts: a faster-expanding outer cone, mainly composed of $\hat{n}_i^{(1)}$, and an accompanying slower inner cone dominated by $\hat{n}_i^{(2)}$~(Fig.~\ref{Fig:1}(a1)). This inner cone reflects the suppressed tunneling rate of the repulsively bound pair of bosons (a doublon)~\cite{wiater2017two}. Consequently, the inner cone is significantly more prominent for the $\vert \psi_{\rm ini}^{(2)} \rangle$ initial state~(Fig.~\ref{Fig:1}(b1)).

We gain further insights into the two-particle QW by noting that the interplay of particle statistics and interactions can be examined by studying the two-particle correlator:
\begin{equation}
    \Gamma_{i,j} (t)  = \langle \psi(t) \vert \hat{a}_i^\dagger \hat{a}_j^\dagger \hat{a}_i \hat{a}_j \vert \psi(t)\rangle.
    \label{eq:Correlator}
\end{equation}
A careful analysis of $\Gamma_{i,j}$ for the $\vert \psi_{\rm ini}^{(1)} \rangle$ reveals that weakly interacting bosons exhibit a characteristic bosonic bunching due to Hanbury Brown–Twiss (HBT) interference. However, as the interaction strength increases, the energy cost of double occupancies leads to the development of anti-bunching correlations akin to those of non-interacting fermions~\cite{lahini2012quantum,preiss2015strongly}. These results are shown in Fig.~\ref{Fig:1}(a2). For the $\vert \psi_{\rm ini}^{(2)} \rangle$ initial state, on the other hand, interactions induce doublon formation, and the density exhibits greater localization with increasing $U$.

These results can be understood by examining the two-particle spectrum of this system~\cite{lahini2012quantum}. For large interaction strengths ($U/J \gg 1$), the spectrum splits into two separate bands: (a) a lower band composed of $L(L-1)/2$ scattering states and (b) an upper band composed of $L$ doublon bound states. The $\vert \psi_{\rm ini}^{(1)}\rangle$ and $\vert \psi_{\rm ini}^{(2)} \rangle$ states are primarily composed of eigenstates from the lower and upper bands, respectively. 

Thus, when $U/J \gg 1$, the time-evolution of the $\vert \psi_{\rm ini}^{(1)}$ state is restricted to a truncated Hilbert space composed only of the scattering states. This in turn implies that in this regime, the bosons become hard-core and the creation and annihilation operators can be described by spin-$1/2$ operators ($\hat{a}_{j} \rightarrow \hat{S}_{j}^{-}$, $\hat{a}_{j}^{\dagger} \rightarrow \hat{S}_{j}^{+}$). We can now employ the Jordan-Wigner transformation to map spin operators to fermionic operators: $\hat{S}_{j}^{-} = e^{- i \hat\theta_{j}} \hat{f}_{j}$ and $\hat{S}_{j}^{+} = e^{ i \hat\theta_{j}} \hat{f}_{j}^{\dagger}$, where the string operator $\hat\theta_{j} = \pi \sum_{l<j} \hat{f}_{l}^{\dagger} \hat{f}_{l}$. We now note that when $\Gamma_{i,j}$ is computed, the string operators cancel, leading to the same value of $\Gamma_{i,j}$ for hard-core bosons and non-interacting fermions.

This concludes our analysis of the two-particle QW in the absence of a dc field. Next, we perform an analogous analysis for the two-particle QW in the presence of a dc field.\\
 
\paragraph{Hermitian quantum walks in the presence of a dc field ($F \neq 0$):}  In the absence of interactions, a dc field leads to Wannier-Stark localization~\cite{wannier1959elements,emin1987existence} and Bloch oscillations~\cite{zener1934theory,anderson1998macroscopic,geiger2018observation,li2022bose}. However, the oscillation pattern changes significantly when $U \ne 0$~\cite{preiss2015strongly,wiater2017two}. Analogous to the $F=0$ case, the density profile fragments into two parts when $U \ne 0$: $\hat{n}_i^{(1)}$ oscillates with the single-particle Bloch oscillation period, $T_B = 2 \pi/F$, while $\hat{n}_i^{(2)}$ oscillates with a period of $T_{B}/2$, leading to the formation of a density hourglass structure~\cite{wiater2017two}. These results are shown in Fig.~\ref{Fig:1}(a3, b3). Analogous to the $F=0$ scenario, the inner hourglass structure is significantly more prominent for the $\vert \psi^{(2)}_{\rm ini} \rangle$ initial state. Finally, we note that recent studies have proposed that these QWs can be employed for quantum-enhanced sensing of weak dc fields~\cite{cai2021multiparticle,yang2024controllable}.\\

\begin{figure}
    \centering
    \includegraphics[width=\linewidth]{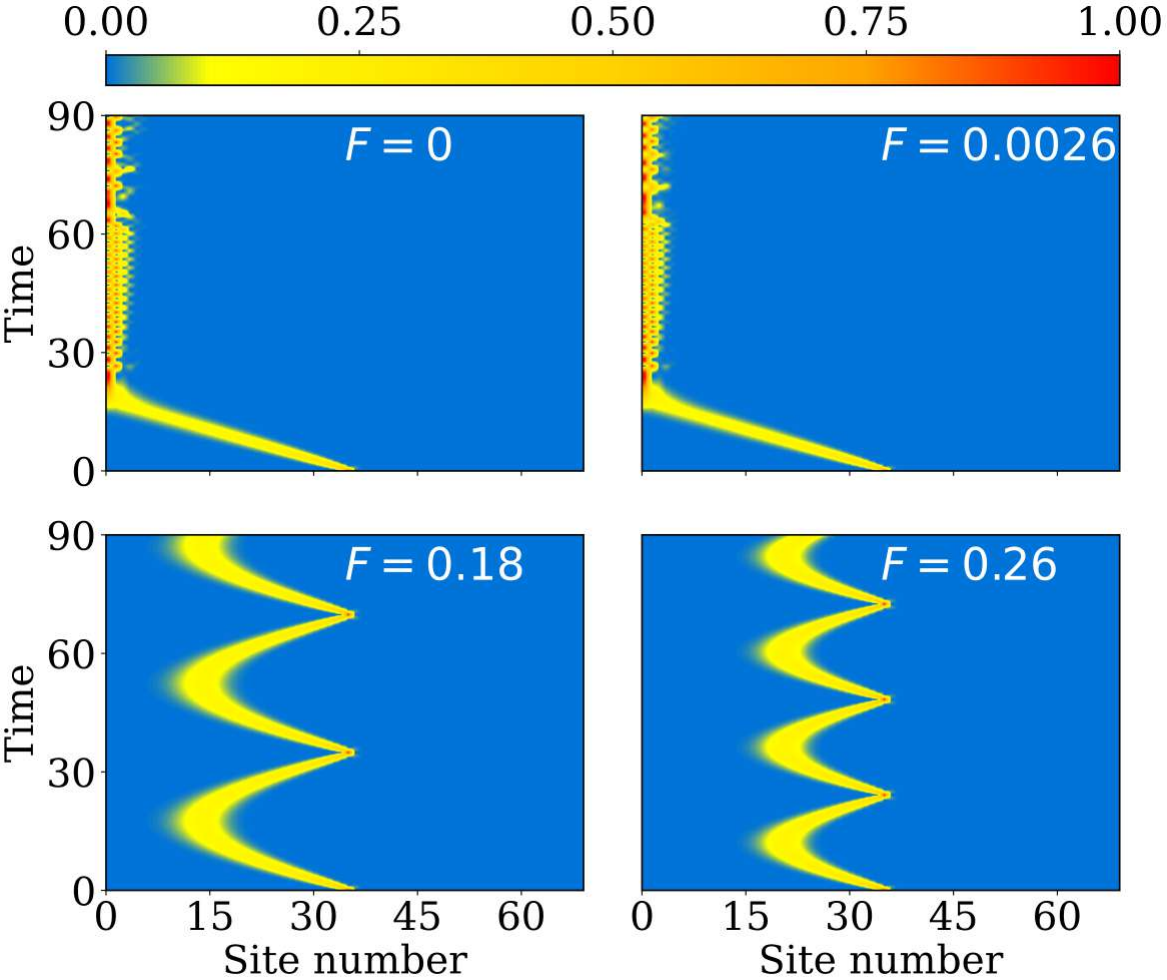}
    \caption{\textbf{Single-particle non-Hermitian QW for $\delta= 0.6$:} The time evolution of the density, $\langle \hat{a}_i^{\dagger} \hat{a}_i \rangle$ for the $\vert \psi_{\rm ini} \rangle = a^{\dagger}_{L/2} \vert 0 \rangle$ and various values of the dc field strength, $F$ at $L=70$. Increasing the dc field strength leads to a crossover from the NHSE to directional Bloch oscillations.}
    \label{Fig:2}
\end{figure}

\paragraph{Non-Hermitian single-particle quantum walks:} Before concluding this section, we review single-particle QWs in the non-Hermitian regime ($\delta \ne 0$). As discussed in Sec.~\ref{sec:Intro}, a striking feature of non-Hermitian systems with open boundary conditions is the NHSE, which is dynamically manifested by directional QWs. Interestingly, the NHSE can be manipulated by a dc field~\cite{peng2022manipulating}. Analogous to the Hermitian case, a dc field leads to Stark localization, and the Bloch oscillations acquire a directionality due to the non-reciprocal nature of the tunneling. Crucially, a competition between Bloch oscillations and the NHSE emerges when the system size $L$ is finite (Fig.~\ref{Fig:2}). To see this, we note that for a particle initially localized at site $n$, the density at time $t$ is given by:
\begin{equation}
\langle \hat{a}^{\dagger}_{j}\hat{a}_{j} \rangle \propto
\mathcal{J}^{2}_{j-n}\!\left(\frac{4J\sqrt{1-\delta^{2}}}{F}
\sin(Ft/2)\right)\left(\frac{1-\delta}{1+\delta}\right)^{j-n},
\end{equation}
where $\mathcal{J}_{j-n}$ is a Bessel function of the first kind with order $j-n$, and the proportionality constant is fixed by the normalization $\langle\sum_j  a_j^{\dagger} a_j\rangle = 1$. Consequently, the system exhibits Bloch oscillations with a range given by $A_{\rm BO} = 4 J(1+\delta)/(Fx_{\ast})$, where $x_{\ast}$ only depends on $J$ and $\delta$~\cite{peng2022manipulating}. Thus, a particle initially localized on site $n$, $\vert \psi_{\rm ini}\rangle = \hat{a}^{\dagger}_{n} \vert 0 \rangle$, would not reach the boundary if $A_{\rm BO} < n$, leading to Bloch oscillations with a period $T_B$. Otherwise, it would stay localized at the boundary due to the NHSE. Consequently, for large $L$, the number of skin modes, $N_{\rm skin}$ is $L-$independent ($N_{\rm skin}\approx A_{\rm BO}$), and the fraction of skin modes in the thermodynamic limit $\lim_{L \rightarrow \infty} N_{\rm skin}/L$ is negligible. This is in sharp contrast to the $F=0$ case, where $\lim_{L \rightarrow \infty} N_{\rm skin}/L$ remains finite, leading to the NHSE. We thus conclude that a nonzero dc field suppresses the NHSE in the thermodynamic limit.\\

So far, we have briefly reviewed some of the key results on both single-particle and two-particle Hermitian QWs in the BH model, as well as single-particle non-Hermitian QWs in the HN model. In the rest of this work, we examine two-particle QWs both in the absence and presence of a dc field in the HNBH model. 

\begin{figure*}[t] 
    \centering
    \includegraphics[width=0.99\textwidth]{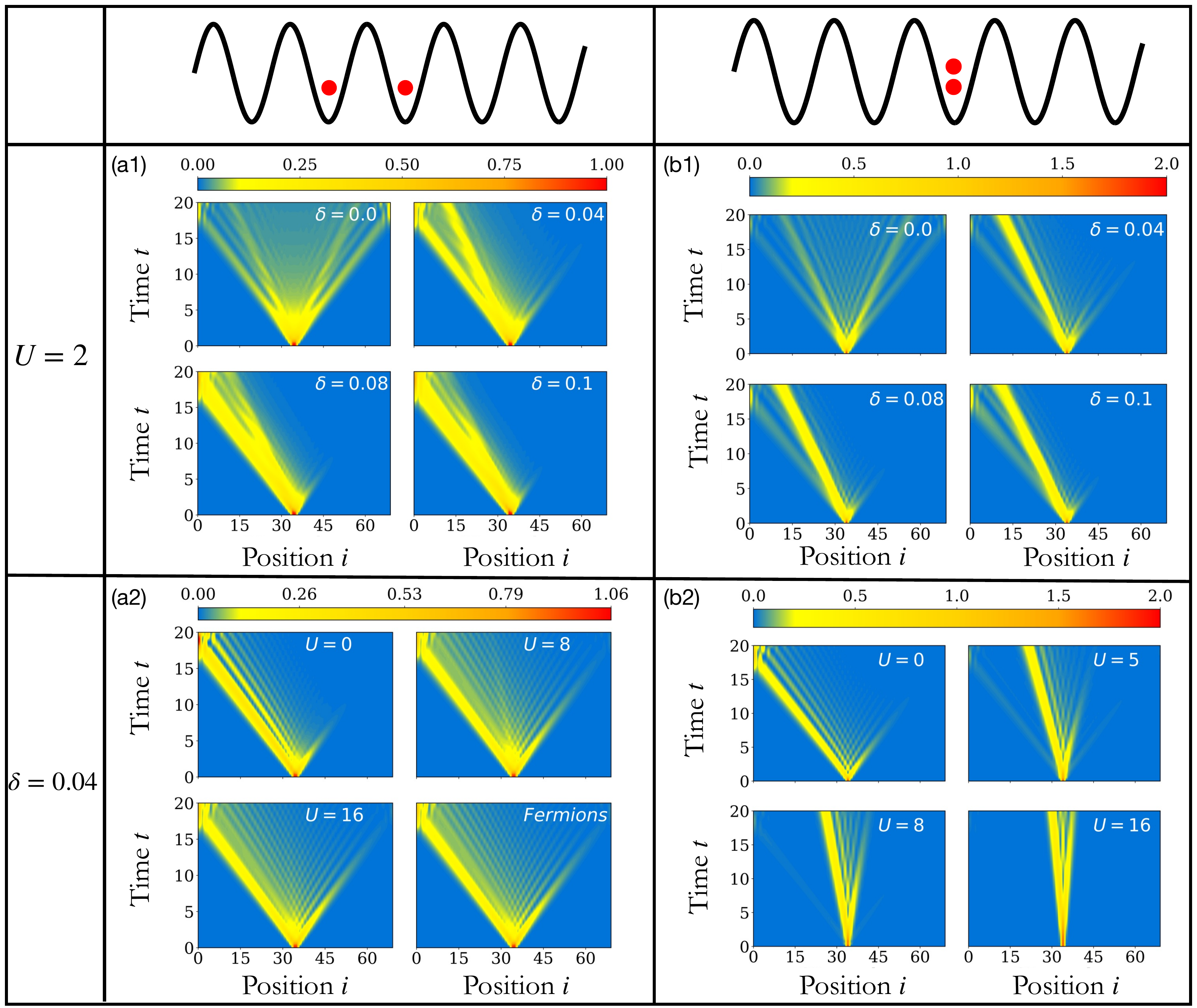}
    \caption{\textbf{The time-evolution of the density distribution, $\hat{n}_i$ (Eq.~\ref{eq:Density}) when $F=0$:} For a fixed interaction strength, $U=2$, $\delta$ induces a transition from symmetric to asymmetric density spreading. This behavior is observed when (a1) the two bosons are initially placed on nearest-neighbor sites, $\vert \psi_{\rm ini}^{(1)} \rangle$ (defined in Eq.~\ref{eq:IniState1})  and (b1) both bosons are initially localized on the same site, $\vert \psi_{\rm ini}^{(2)} \rangle$ (defined in Eq.~\ref{eq:IniState2}). A complementary analysis for fixed $\delta = 0.04$ reveals that strong interactions induce effective fermionization for $\vert \psi_{\rm ini}^{(1)} \rangle$ and doublon localization for $\vert \psi_{\rm ini}^{(2)} \rangle$. } 
    \label{Fig:3}
\end{figure*}
\begin{figure*}[t] 
    \centering
    \includegraphics[width=0.99\textwidth]{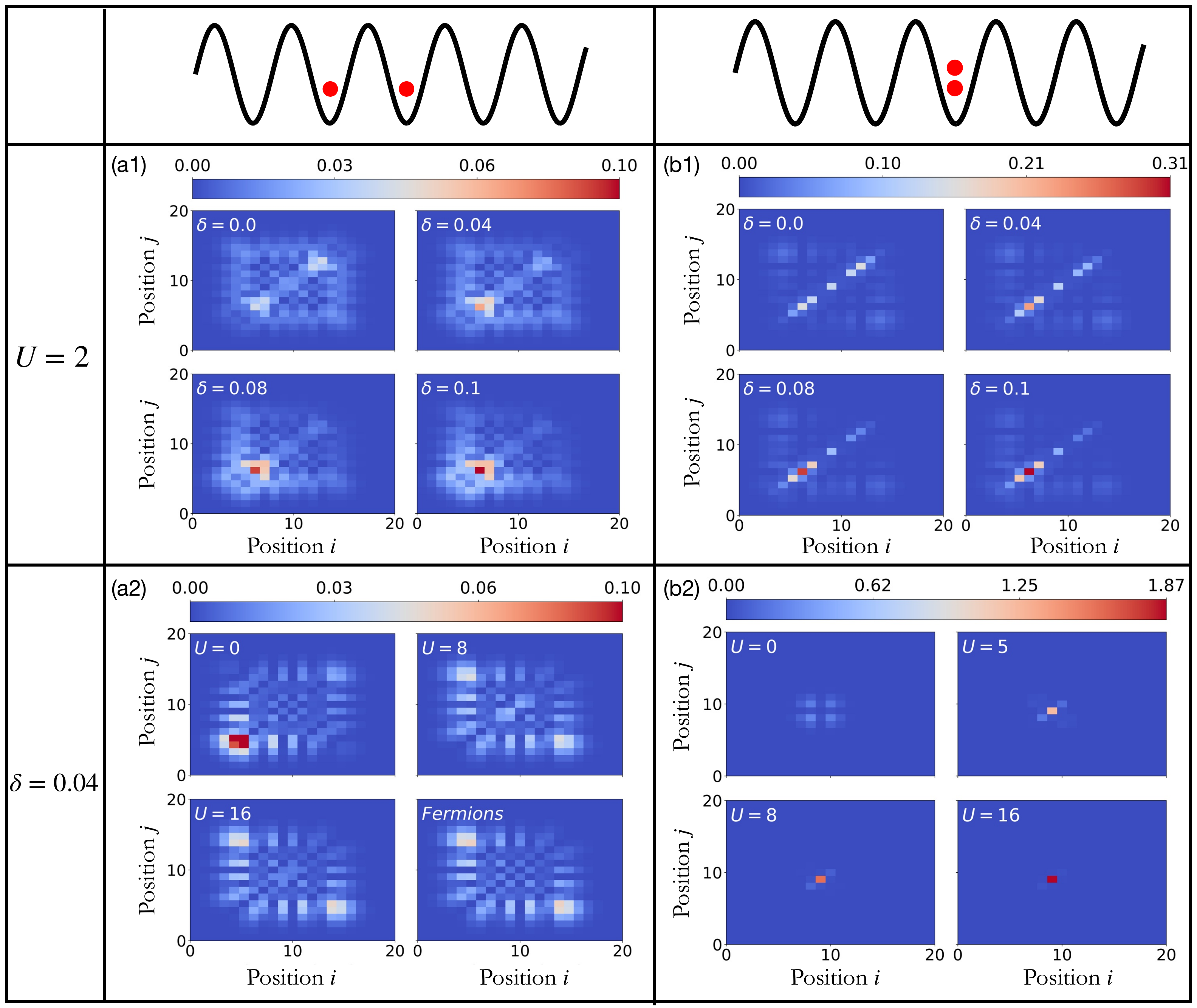}
    \caption{\textbf{The two-particle correlator $\Gamma_{i,j}$ (Eq.~\ref{eq:Correlator}) in the absence of a dc field ($F=0$):} For fixed $U=2$, increasing $\delta$ induces stronger non-reciprocal bunching for both initial states: (a1) $\vert \psi_{\rm ini}^{(1)} \rangle$ (Eq.~\ref{eq:IniState1}) and (b1) $\vert \psi_{\rm ini}^{(2)} \rangle$ (Eq.~\ref{eq:IniState2}). On the other hand, for fixed $\delta =0.04$, large $U$ induces an effective fermionization and consequent non-reciprocal anti-bunching for $\vert \psi_{\rm ini}^{(1)} \rangle$ and doublon localization for $\vert \psi_{\rm ini}^{(2)}\rangle$. $\Gamma_{i,j}$ has been computed at $t=\pi$ for panels (a1), (a2) and (b1), $t=1$ for panel (b2).} 
    \label{Fig:4}
\end{figure*}
\section{Non-Hermitian Quantum Walks in the absence of a dc field}
\label{sec:NHSE}
In this section, we focus on the QW of two bosons in the absence of an external dc field ($F=0$) in the presence of non-reciprocal tunneling ($\delta \ne 0$). To this end, we first explore the dynamics of the system for a fixed interaction strength, $U$ and varying the non-reciprocity parameter, $\delta$. Our results are shown in Fig.~\ref{Fig:3}(a1) (for the $\vert \psi_{\rm ini}^{(1)} \rangle$ initial state) and Fig.~\ref{Fig:3}(b1) (for the $\vert \psi_{\rm ini}^{(2)} \rangle$ initial state); we set $U=2$ and $L=70$ for these calculations. As discussed in Sec.~\ref{sec:Model}, in the Hermitian limit ($\delta=0$), the density exhibits ballistic spreading, with a fragmentation into two cones originating from the interactions. However, as $\delta$ increases, the density cone becomes increasingly asymmetric, highlighting the directional nature of the quantum walk. Next, we explore the complementary situation: where  $\delta$ is fixed, and $U$ is varied. We present our results in Fig.~\ref{Fig:3}(a2) for $\vert \psi_{\rm ini}^{(1)} \rangle$ and Fig.~\ref{Fig:3}(b2) for $\vert \psi_{\rm ini}^{(2)} \rangle$. For the $\vert \psi_{\rm ini}^{(1)} \rangle$ initial state, the system `fermionizes' at large $U$ and the density cone is analogous to that of non-interacting fermions. In contrast, for the $\vert \psi_{\rm ini}^{(2)} \rangle$ initial state, the system exhibits a slower spread as $U$ increases due to the suppression of doublon mobility at strong interactions.

To gain further insights into the dynamics of the system for the $\vert \psi_{\rm ini}^{(2)} \rangle$ initial state, we employ perturbation theory (described in App.~\ref{App:A}) to obtain an effective Hamiltonian that describes the doublon dynamics in the $J/U \ll 1$ regime (up to additional constant terms):
\begin{equation}
H_{\mathrm{eff}}
=
J_R \sum_{j=0}^{L-1} \hat{b}_{j+1}^{\dagger}\hat{b}_j
+
J_L \sum_{j=0}^{L-1} \hat{b}_j^{\dagger}\hat{b}_{j+1},
\label{eq:Heff}
\end{equation}
where $J_{R} = \dfrac{2J^2}{U}(1-\delta)^2$ and $J_{L} = \dfrac{2J^2}{U}(1+\delta)^2$, and $\hat{b}_{j} (\hat{b}_{j}^{\dagger})$ is the doublon annihilation (creation) operator at site $j$. Thus, the doublon QW is effectively described by a single-particle HN Hamiltonian. Notably, the effective non-reciprocity in this system, quantified by $\frac{J_L-J_R}{J_L + J_R}= \frac{2 \delta}{1+\delta^2}$, is independent of $U$. Consequently, the directionality of the quantum walk is controlled by $\delta$ alone. However, the average tunneling energy scale is controlled by both $U$ and $\delta$, since $(J_L + J_R)/2 =  \dfrac{2J^2}{U} (1+ \delta^2)$. Taken together, these results elucidate the origin of the directional walk of the doublons, and the enhanced localization at large $U$.

Before proceeding further, it is instructive to discuss the validity of the effective description in Eq.~(\ref{eq:Heff}). Analogous to the
Hermitian case discussed in Sec.~\ref{sec:Model}, the two-particle spectrum
splits into two bands when $U/J \gg 1$: a lower band of scattering states and an upper band of doublon bound states. For the effective description to hold, the doublon band must be spectrally separated from the scattering band. As discussed in App.~\ref{App:C}, this requires $U > 4J\sqrt{1-\delta^{2}}$. Furthermore, the doublon occupation probability, $P_{\rm doub} = \frac{1}{2}\sum_{i} \hat{n}_{i}^{(2)}$, must remain close to $1$ during the time evolution. At finite $U/J$, two small corrections reduce $P_{\rm doub}$: (a) the initial state $\lvert \psi^{(2)}_{\rm ini} \rangle$ has a small overlap with the eigenstates of the scattering band, and (b) the doublon band eigenstates carry a small weight on configurations in which the two bosons occupy neighboring sites. Consequently, $P_{\rm doub}$ settles to a plateau at
$P_{\rm doub} \simeq 1 - 8J^{2}/U^{2}$, which is independent of $\delta$ up to corrections of order $\delta^{2}$. As shown in App.~\ref{App:C}, this analytical prediction agrees closely with our numerical results: after a short transient, $P_{\rm doub}$ settles at $\approx 0.97$ ($\approx 0.89$) for $U = 16$ ($8$), for both $\delta = 0$ and $\delta = 0.04$. The effective description is therefore quantitatively accurate for $U \gtrsim 8$, and remains
qualitatively reliable even for $U = 5$.
\begin{figure*}[t] 
    \centering
    \includegraphics[width=0.99\textwidth]{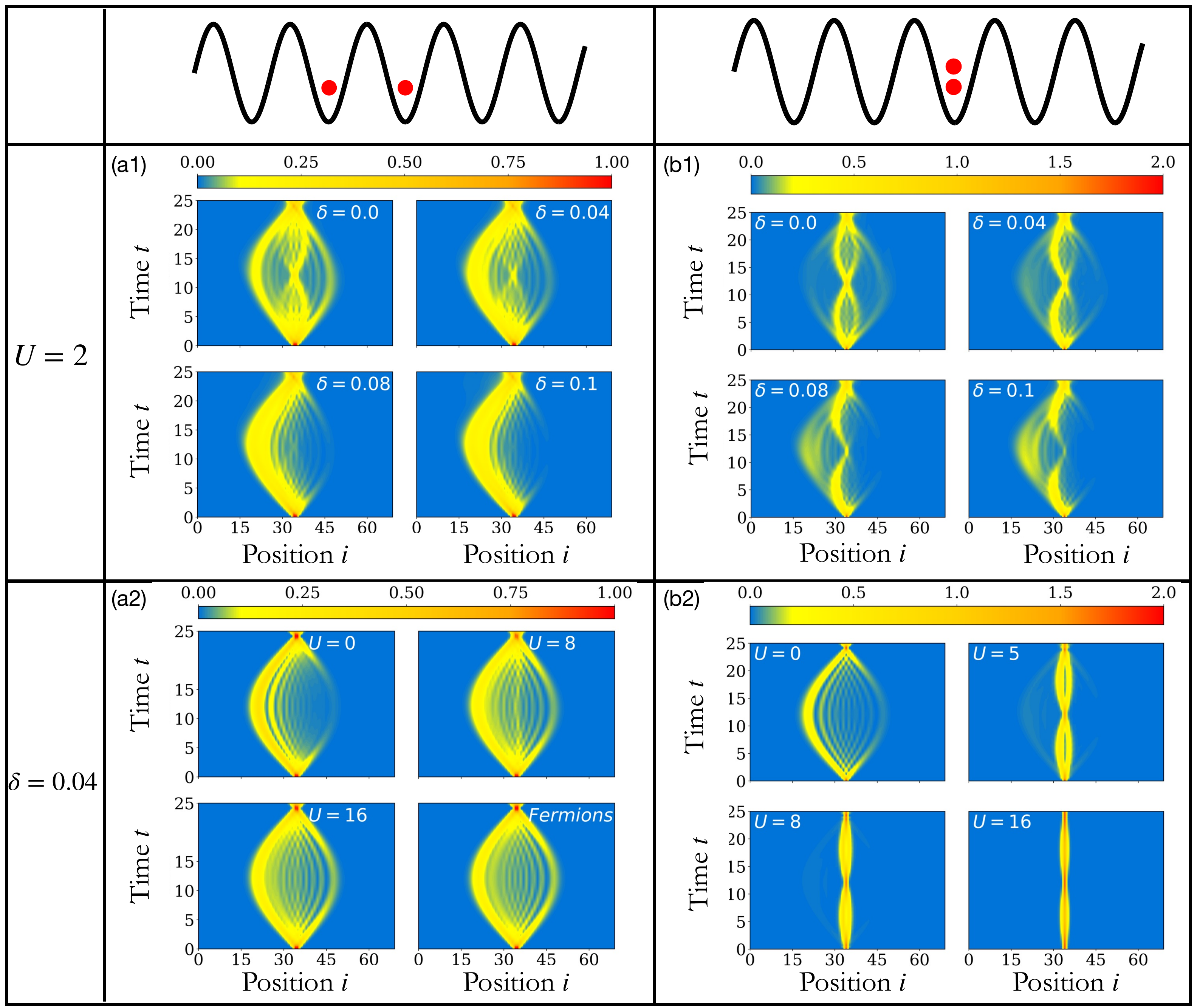}
    \caption{\textbf{Time evolution of the density in the presence of a dc field ($F=0.26$):} For a fixed interaction strength $U=2$, co-walking dynamics give rise to an hourglass-shaped structure for both initial states: (a1) $\vert \psi_{\rm ini}^{(1)} \rangle $ (Eq.~\ref{eq:IniState1}) and (b1) $\vert \psi_{\rm ini}^{(2)} \rangle$ (Eq.~\ref{eq:IniState2}). As the non-reciprocity parameter $\delta$ is increased, asymmetric Bloch oscillations emerge, accompanied by a gradual suppression of the hourglass structure for $\vert \psi_{\rm ini}^{(1)} \rangle$, and a concomitant enhancement for $\vert \psi_{\rm ini}^{(2)} \rangle$. On the other hand, for fixed $\delta = 0.04$, increasing $U$ leads to fermionization and consequently more symmetric Bloch oscillations for $\vert \psi_{\rm ini}^{(1)} \rangle$. In contrast, for $\vert \psi_{\rm ini}^{(2)} \rangle$, larger $U$ enhances the contribution of the doublon Bloch oscillations to the QW.} 
    \label{Fig:5}
\end{figure*}

\begin{figure*}[t] 
    \centering
    \includegraphics[width=0.99\textwidth]{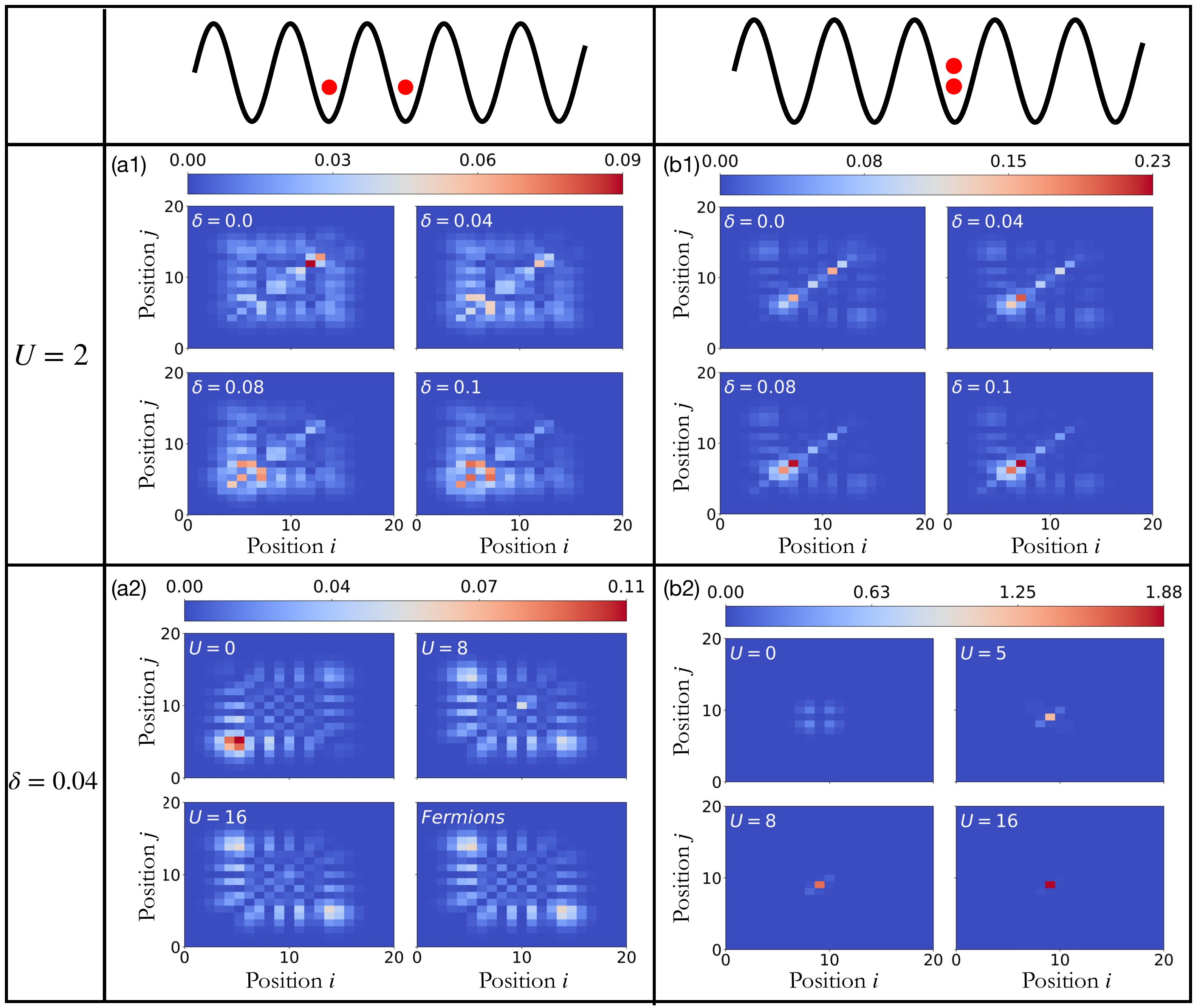}
    \caption{\textbf{The two-particle correlator $\Gamma_{i,j}$ (Eq.~\ref{eq:Correlator}) in the presence of a dc field ($F=0.26)$:} Consistent with the $F=0$ case (Fig.~\ref{Fig:4}), increasing $\delta$ at a fixed $U=2$, leads to non-reciprocal bunching for both (a1) $\vert \psi_{\rm ini}^{(1)} \rangle$ and (b1) $\vert \psi_{\rm ini}^{(2)} \rangle$. For fixed $\delta=0.04$, large $U$ leads to non-reciprocal anti-bunching for $\vert \psi_{\rm ini}^{(1)} \rangle$ and doublon formation for $\vert \psi_{\rm ini}^{(2)} \rangle$. All the other parameters are the same as those of (Fig.~\ref{Fig:4}).} 
    \label{Fig:6}
\end{figure*}

We now proceed to examine the correlations between the two bosons for both the $\vert \psi_{\rm ini}^{(1)} \rangle$ (Fig.~\ref{Fig:4}(a1)-(a2)) and $\vert \psi_{\rm ini}^{(2)} \rangle$ states (Fig.~\ref{Fig:4}(b1)-(b2)). We first examine the correlator, $\Gamma_{i,j}$ (defined in Eq.~\ref{eq:Correlator}) when $U=2$ for different values of $\delta$. We find that in the Hermitian limit ($\delta = 0$), the weight of $\Gamma_{i,j}$ is symmetrically concentrated along the diagonal for both initial states. When $\delta$ is turned on, however, the distribution becomes asymmetric, revealing non-reciprocal bunching. Furthermore, we analyze $\Gamma_{i,j}$, when $\delta$ is fixed ($\delta = 0.04$) and $U$ is changed. When $U=0$, the bosons exhibit non-reciprocal bunching correlations due to the interplay of HBT interference and non-Hermiticity. For $\vert \psi_{\rm ini}^{(1)} \rangle$, as $U$ increases, the weight of $\Gamma_{i,j}$ gradually shifts towards the anti-diagonal, indicating a non-reciprocal fermion-like anti-bunching behavior. On the other hand, for $\vert \psi_{\rm ini}^{(2)} \rangle$, increasing $U$ leads to greater doublon localization. 

This analysis concludes our discussion of two-particle bosonic QWs in the absence of a dc field ($F=0$). In this section, we have established that the non-reciprocal tunneling in the HNBH model leads to directional quantum walks and the consequent formation of an asymmetric density cone. This is accompanied by the development of non-reciprocal bunching (anti-bunching) correlations in the presence of weak (strong) interactions.  We now proceed to investigate the capabilities of non-Hermitian QWs of two bosons in the presence of a dc field.

\section{Non-Hermitian Quantum Walks in the Presence of a dc Field}
\label{sec:Bloch}

In this section, we examine the QW of two bosons in the presence of a finite dc field, $F$, for both the $\vert \psi_{\rm ini}^{(1)} \rangle$ and $\vert \psi_{\rm ini}^{(2)} \rangle$ initial states. As discussed in Sec.~\ref{sec:Model}, $F$ induces a Wannier-Stark ladder in the single-particle energy spectrum, leading to Bloch oscillations with frequency $\omega = F$ when $U=0$. These oscillations become asymmetric in the presence of non-Hermiticity.

We now examine the QW of two bosons when $U \ne 0$; we set $F=0.26$ for our calculations. Analogous to the analysis in Sec.~\ref{sec:NHSE}, we first analyze the QW when the interaction strength, $U$ is fixed ($U=2$), and $\delta$ is varied. Our results are shown in Fig.~\ref{Fig:5}(a1) (for $\vert \psi_{\rm ini}^{(1)} \rangle$) and Fig.~\ref{Fig:5}(b1) (for $\vert \psi_{\rm ini}^{(2)} \rangle$). In both cases, for $\delta=0$, an hourglass-like structure appears within the Bloch oscillations, originating from two-particle co-walking, whose oscillation period is half of the Bloch oscillation period. This hourglass structure is more prominent for $\vert \psi_{\rm ini}^{(2)} \rangle$. When $\delta \ne 0$, the oscillations become directional, similar to the single-particle walk. However, there is a strong initial state-dependence of the hourglass structure. For $\vert \psi_{\rm ini}^{(1)} \rangle$, the hourglass structure gradually diminishes with increasing $\delta$. However, for $\vert \psi_{\rm ini}^{(2)} \rangle$, the hourglass structure is retained, and it becomes increasingly asymmetric with increasing $\delta$.

Next, we explore the complementary scenario, by fixing $\delta$ ($\delta = 0.04$) and varying $U$. Our results are shown in Fig.~\ref{Fig:5}(a2) (for $\vert \psi_{\rm ini}^{(1)} \rangle$) and Fig.~\ref{Fig:5}(b2) (for $\vert \psi_{\rm ini}^{(2)} \rangle$). We find that in both cases, the non-interacting quantum walk is strongly asymmetric, in accordance with known results from the single-particle QW~\cite{peng2022manipulating}. Interestingly, for the $\vert \psi_{\rm ini}^{(1)} \rangle$ initial state, the oscillations become increasingly more symmetric with increasing $U$. This behavior originates from the suppression of double occupancy at large $U$, leading to QW dynamics analogous to those of non-interacting fermions. On the other hand, for the $\vert \psi_{\rm ini}^{(2)} \rangle$ initial state, the nature of the walk is primarily dictated by the doublons at large $U$. This leads to the emergence of correlated Bloch oscillations with a period $T_B/2$. These oscillations are more symmetric than their non-interacting counterpart and the doublons become localized at large $U$.

Analogous to the $F=0$ case, the effective doublon Hamiltonian that describes the dynamics of the $\vert \psi^{(2)}_{\rm ini}\rangle$ state in the $J/U \ll 1$ regime (up to additional constant terms)  is given by (see App.~\ref{App:B}) :
\begin{equation}
\hat{H}_{\mathrm{eff}} = J_{R} \sum_{j=0}^{L-1} \hat{b}_{j+1}^{\dagger}\hat{b}_j + J_{L}
\sum_{j=0}^{L-1} \hat{b}_{j}^{\dagger}\hat{b}_{j+1} + 2F \sum_{j=0}^L j \hat{n}_j ,
\end{equation} 
where $J_{R} = \frac{2J^2}{U}(1-\delta)^2$, 
$J_{L} = \frac{2J^2}{U}(1+\delta)^2$, and the doublon is initially localized at $j=L/2$, i.e., $\vert \psi (t=0)\rangle = \hat{b}_{L/2}^{\dagger} \vert 0 \rangle$. Thus, the doublon QW maps onto the single particle QW in the HN model in the presence of a dc field with magnitude, $2F$. We note that the effective description presented here is valid under the same condition as the $F = 0$ case discussed in Sec.~\ref{sec:NHSE}: $U > 4J\sqrt{1-\delta^{2}}$. A straightforward way to understand this is to note that the dc field reduces to a uniform time-dependent phase on the tunneling amplitudes in the rotating frame presented in App.~\ref{App:B}. This phase can be removed by a gauge
transformation at any instant on an open chain. Consequently, the instantaneous spectrum, and hence the separation between the doublon and scattering bands, is identical to that at $F = 0$. Consistent with this analysis, the plateau value of $P_{\rm doub}$ is not appreciably changed by the dc field considered here (see App.~\ref{App:C}). Furthermore, it is worth pointing out that a resonant dissociation of the doublon can occur for fine-tuned values of the interaction strength, $U \simeq dF$ with $d$ a positive integer, leading to tunneling oscillations~\cite{khomeriki2010interaction}. A sharp signature of this effect, however, is expected only in the strong-field regime in which the doublon Bloch oscillations are confined to a single lattice site~\cite{khomeriki2010interaction}, which lies outside the parameter range considered here.

We now proceed to employ the results stated in Sec.~\ref{sec:Model}(c) to gain further insights into the doublon QW. In this case, the doublon  density at site $j$ is given by: 
\begin{eqnarray}
    \langle \hat{b}_j^{\dagger}\hat{b}_j \rangle &=& \mathcal{N} \mathcal{J}^2_{j-L/2}\bigg(2 \sqrt{J_L J_R}/F \sin(Ft) \bigg) (J_R/J_L)^{j-L/2} \nonumber\\
    &=&\mathcal{N}\mathcal{J}^2_{j-L/2}\bigg(\frac{4J^2 (1-\delta^2)}{U F} \sin(Ft) \bigg) \bigg(\frac{1-\delta}{1+\delta}\bigg)^{2j-L},\nonumber \\
\end{eqnarray}
where $\mathcal{J}_{j-L/2}$ is a Bessel function of the first kind with order $j-L/2$ and $\mathcal{N}$ is a time-dependent constant fixed by $\langle \sum_j b_j^{\dagger} b_j \rangle = 1$~\cite{peng2022manipulating}. For the $\delta$ values considered in this work ($\delta \le 0.1$), the argument of the Bessel function does not exhibit any appreciable $\delta$ dependence. Employing the fact that $\sin(Ft) = 0$ at $t = n \pi/F\ \forall n \in \mathbb{Z}$, $\mathcal{J}_{j-L/2} (0) = \delta_{j-L/2,0}$, we obtain that $\langle \hat{b}_{L/2}^{\dagger}\hat{b}_{L/2} \rangle (t= n \pi/F) =1$. This implies that the doublon exhibits BO with a period, $T_B^{d} = T_B/2 = \frac{\pi}{F}$. We note that the factor $(J_R/J_L)^{j-L/2}$ induces the directionality in the BO as evidenced in Fig.~\ref{Fig:5}. Finally, we note that the amplitude of the doublon BO, $A_{\rm BO}^d$, takes the same form as $A_{\rm BO}$ discussed in  Sec.~\ref{sec:Model}(c), namely $A_{\rm BO}^d = 4 J^2(1+\delta)^2/(U Fx_{\ast}^{d})$, where $x_{\ast}^{d}$ is the doublon analog of $x_{\ast}$ and depends only on $J_L$ and $J_R$. 

We conclude our discussion by analyzing the two-particle correlator $\Gamma_{i,j}$ (Eq.~\ref{eq:Correlator}), for the two bosons in the presence of a dc field. We find that the non-Hermiticity induces non-reciprocal bunching of the bosons at weak $U$ and a non-reciprocal anti-bunching at strong $U$ for the $\vert \psi_{\rm ini}^{(1)} \rangle$ initial state (see Fig.~\ref{Fig:6}(a2)). On the other hand for the $\vert \psi_{\rm ini}^{(2)} \rangle$ initial state, increasing $U$ leads to greater doublon localization (see Fig.~\ref{Fig:6}(b2)). This behavior mirrors the results obtained for the $F=0$ case discussed in Sec.~\ref{sec:NHSE}.\\

\begin{figure*}
    \centering
    \includegraphics[width=\textwidth]{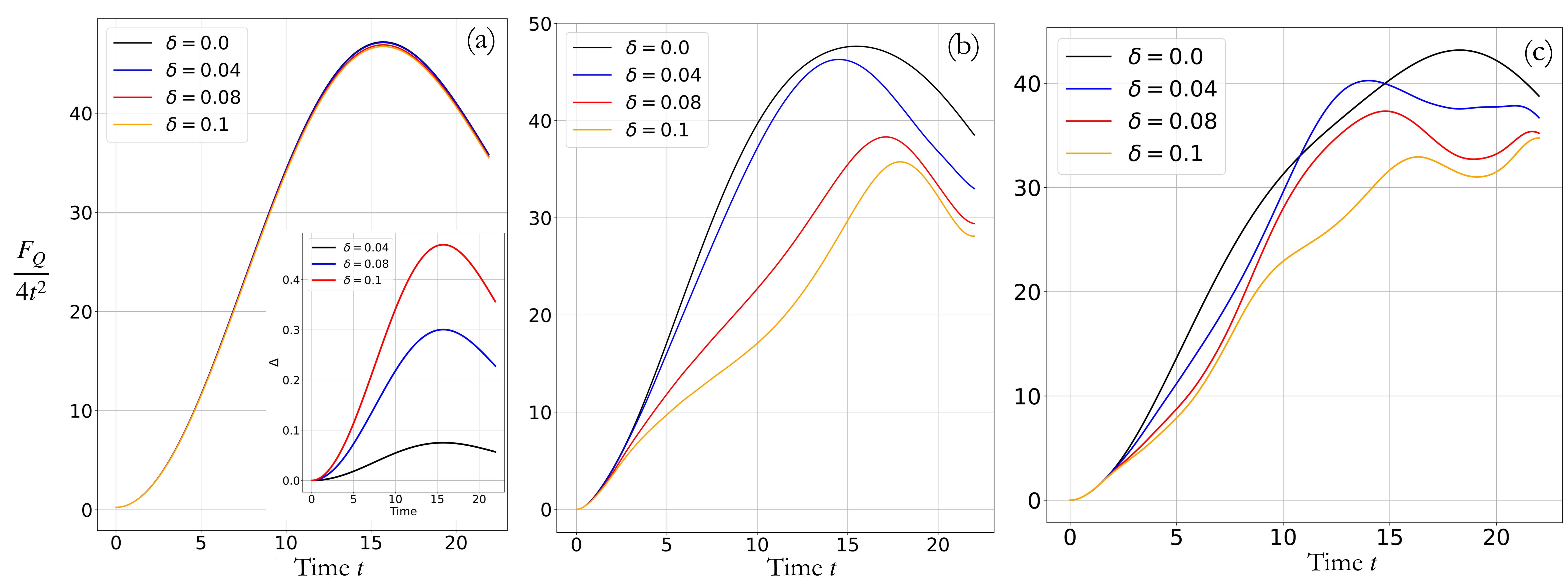}
    \caption{\textbf{The time evolution of the Quantum Fisher Information, $F_Q$ (Eq.~\ref{eq:QFI}) for single- and two-particle QWs at different values of $\delta$ for $F=0.26$:} (a) For the single-particle QW, $F_Q \propto t^{\alpha}$, where $\alpha \simeq 3$ for $t \lesssim T_B/2$ both when $\delta=0$ and $\delta \ne 0$. The inset shows the slow growth of $\Delta$ (Eq.~\ref{eq:QFIdiff}) with time, thereby establishing that $\delta$ induces a minimal decrease in $F_Q$. For the two-particle QWs, the time-evolution of $F_Q$ for $\vert \psi_{\rm ini}^{(1)} \rangle$ (b) and $\vert \psi_{\rm ini}^{(2)} \rangle$ (c) shows that although increasing $\delta$ suppresses $F_Q$, a rapid temporal growth persists, with $F_Q \propto t^{\alpha}$, where $\alpha \simeq 3$. } 
    \label{Fig:7}
\end{figure*}

In this section, we have established that in the presence of a dc field, the non-reciprocity of the HNBH model leads to asymmetric Bloch oscillations. Notably, strong interactions tend to reduce the degree of asymmetry in these oscillations. Moreover, when both bosons are initially localized on the same site, co-walking dynamics lead to the emergence of an asymmetric hourglass-shaped density profile. Finally, we have demonstrated that, analogous to the $F=0$ case, this system exhibits non-reciprocal bunching and anti-bunching in the weak and strong interaction regimes, respectively. We now proceed to the final part of this work, where we examine the growth of the quantum Fisher information for the non-Hermitian QWs.

\section{Quantum Fisher information dynamics}
\label{sec:QFI}

As noted in Sec.~\ref{sec:Model}, recent studies have proposed that Hermitian QWs can be harnessed for sensing weak dc forces, $F$~\cite{cai2021multiparticle,yang2024controllable}. This sensitivity is characterized by the quantum Fisher information (QFI), $F_Q$~\cite{liu2020quantum,braunstein1994statistical}:
\begin{equation}
{F_Q} = 4 \left\{\left[ \frac{\partial}{\partial F} \langle \psi(t) | \right] 
\frac{\partial}{\partial F} | \psi(t) \rangle - \left| \langle \psi(t) | \frac{\partial}{\partial F} | \psi(t) \rangle \right|^2 \right\}.
\label{eq:QFI}
\end{equation}
We note that $F_Q$ sets a fundamental bound on the precision with which $F$ can be estimated~\cite{agarwal2025quantum}:
\begin{equation}
    \Delta F \ge 1/\sqrt{F_Q}.
\end{equation}
For the Hermitian case there is a characteristic time $t_{0} \approx 0.5 \, T_{B}$  $(T_{B} = \frac{2\pi}{F})$ below which $F_Q$ grows approximately as $t^3$~\cite{cai2021multiparticle}, thereby providing a route for the sensitive measurement of weak dc fields. We now investigate the fate of $F_Q$ in the non-Hermitian regime ($\delta \ne 0$) for both single and two-particle QWs.\\

Figure~\ref{Fig:7}(a) shows that for the one-particle walk, $F_Q$ continues to grow approximately as $ t^3$ and it decreases very slowly with increasing $\delta$. We further establish the weak dependence of $F_Q$ on $\delta$ by computing
\begin{equation}
\Delta = \left( \frac{F_Q}{4t^{2}} \bigg|_{\delta=0} \right) - \left( \frac{F_Q}{4t^{2}} \bigg|_{\delta} \right),
\label{eq:QFIdiff}
\end{equation}
and demonstrating that $\Delta$ remains very small for finite $\delta$ (see inset of Fig.~\ref{Fig:7}(a)). Next we investigate the time-evolution of $F_Q$ for two-particle QWs for the two initial states: $\vert \psi_{\rm ini}^{(1)} \rangle$ (Fig.~\ref{Fig:7}(b)) and $\vert \psi_{\rm ini}^{(2)} \rangle$ (Fig.~\ref{Fig:7}(c)). It is evident from these figures that increasing $\delta$ suppresses the growth of $F_Q$. This decrease in $F_Q$ reflects the decrease in entanglement entropy induced by non-reciprocal tunneling~\cite{kawabata2023entanglement}. Despite this slower growth, we find that the scaling behavior of $F_Q$ remains almost unchanged with varying $\delta$ ($F_Q \propto t^{\alpha}$, where $\alpha \simeq 3$). \\

We note, however, that the rapid growth of $F_Q$ persisting in the
presence of non-Hermiticity does not by itself constitute a demonstration of quantum-enhanced sensing. The HN model can be realized as the dynamics of a continuously monitored system along the trajectory in which no quantum jump has occurred~\cite{daley2014quantum, kawabata2023entanglement}. In such a realization, the $F_Q$ computed here would quantify the precision attainable conditioned on this no-jump evolution, and a complete figure of merit would have to account for this post-selection. On the other hand, it is worth noting that the HN model in flat space is dual to a Hermitian system on a hyperbolic
surface, with the non-Hermiticity parameter setting the curvature~\cite{lv2022curving}. Synthetic curved spaces therefore
present an alternative route to emulating non-reciprocal tunneling
without monitoring and post-selection. In such a realization the
evolution is unitary, and $F_Q$ directly bounds the attainable
sensitivity. Extending this duality to the HNBH model in the presence of a dc field remains an open question. We conclude that the
appropriate figure of merit for sensing with quantum walks in the HNBH
model depends on the specific realization of the dynamics, and a
detailed analysis of this question is a promising direction for future
research.

\section{Summary and Outlook}
\label{sec:Conclusion}

In this work, we have analyzed the QW of two interacting bosons in a Hatano-Nelson lattice. We have demonstrated that in the absence of a dc field, non-reciprocity induces an asymmetric density cone in the time-evolution of the system. This is accompanied by non-reciprocal bunching correlations in the weak interaction regime, and fermion-like anti-bunching correlations in the strong interaction regime, when the two bosons are placed on neighboring sites. When both bosons are placed on the same site, then the system exhibits localization due to the reduced mobility of the doublon. Next, we have analyzed the two-boson QW in the presence of a dc field and demonstrate that the interplay of interactions, the dc field, and non-reciprocity leads to the formation of an asymmetric hourglass structure in the time-evolution of the density distribution. At strong interactions, the hourglass structure becomes weaker (stronger) when the bosons are initially placed on neighboring sites (same site). Finally, we compute the quantum Fisher information, $F_Q$ and demonstrate that $F_Q \propto t^{\alpha}$, where $\alpha \simeq 3$ for both one- and two-particle QWs, just like their Hermitian counterparts. Thus, our work presents a comprehensive analysis of two-particle QWs for bosons in a Hatano-Nelson lattice.

There are several directions of future research that can possibly stem from this work. A natural next step would be to investigate non-Hermitian multi-particle walks in the presence of random and quasi-periodic disorder. It would also be interesting to examine the interplay of long-range tunneling and non-Hermiticity on these systems. Finally, exploring anyonic quantum walks~\cite{kwan2024realization} in the presence of non-reciprocal tunneling could be an intriguing direction of research.

\section*{Acknowledgments}
SC thanks DST, India, for support through the project DST/FFT/NQM/QSM/2024/3. KS has been supported by the Visiting Students Program at HRI. SA deeply appreciates the enlightening and constructive discussions with Shuva Mondal.\\

\appendix
\section{Effective doublon Hamiltonian in the absence of an external field $(F=0)$}
\label{App:A}
In this appendix, we derive the effective Hamiltonian that describes the doublon QW in the absence of an external field $(F=0)$. To this end, we apply perturbation theory in the $J/U \ll 1$ regime to obtain an effective Hamiltonian of the form:
\begin{equation}
    \hat{H}_{\rm eff} = \hat{H}^{(0)} + \hat{H}^{(1)} + \hat{H}^{(2)}+ \ldots,
\end{equation}
where
\begin{equation}
\hat{H}^{(0)} =  \frac{U}{2} \sum_{j=0}^{L} \, \hat{a}_j^{\dagger} \hat{a}_j^{\dagger} \hat{a}_j \hat{a}_j \nonumber,
\end{equation}
is the unperturbed Hamiltonian, and we treat the tunneling term, $V$ as the perturbation:
\begin{equation}
\hat{V} = \sum_{j=0}^{L-1} -J\left[ (1 - \delta)\, \hat{a}_{j+1}^{\dagger} \hat{a}_j + (1 + \delta)\, \hat{a}_j^{\dagger} \hat{a}_{j+1} \right] \nonumber ,
\end{equation}

In the two-particle sector, the unperturbed Hamiltonian, $H^{(0)}$ has only two eigenvalues: (i) $E_0 = U$ and $\tilde{E}_0 = 0$. The $E_0 = U$ eigenstates correspond to the $L$-fold degenerate states composed of a doubly occupied state at site $j$, $\vert G_j\rangle = \frac{1}{\sqrt{2}} (\hat{a}_{j}^{\dagger})^2 \vert 0\rangle = b_j^{\dagger} \vert 0 \rangle$, where $b_j^{\dagger}$ is the doublon creation operator at site j. On the other hand, the $\tilde{E}_0 = 0$ eigenstates correspond to states of the form: $\vert E_{\rm i,j}\rangle =\hat{a}_{i}^{\dagger} \hat{a}_{j}^{\dagger} \vert 0 \rangle$ for $i \ne j$.

The first-order contribution to the effective Hamiltonian is then given by~\cite{qin2014statistics,yang2024controllable}:
\begin{equation}
    \hat{H}^{(1)} = \hat{P}_{G} \hat{V} \hat{P}_{G},
\end{equation}
where $\hat{P}_{G} = \sum_{j} \vert G_j \rangle \langle G_j \vert$. Then, a straightforward algebra yields
\begin{align*}
{H}^{(1)} = 0.
\end{align*}

It is thus crucial to obtain the effective Hamiltonian to second order in perturbation theory. This is done by:
\begin{equation}
\hat{H}^{(2)} = \hat{P}_{G} \hat{V} \frac{1 - \hat{P}_{G}}{E_{0} - H^{(0)}} \hat{V} \hat{P}_{G}.
\label{eq:perturb}
\end{equation}
Evaluating Eq.~\ref{eq:perturb}, we obtain:
\begin{widetext}
\begin{eqnarray}
\hat{H}^{(2)} &=& \sum_j \bigg[\frac{2J^2}{U} (1+\delta)^2 |G_{j-1} \rangle \langle G_{j}| + \frac{4J^2}{U} (1-\delta^2) |G_{j} \rangle \langle G_{j}| + \frac{2J^2}{U} (1-\delta)^2 |G_{j+1} \rangle \langle G_{j}|\bigg] \nonumber \\
&=& \frac{4J^2}{U} (1-\delta^2) \sum_{j} \hat{b}_{j}^{\dagger} \hat{b}_{j} + \frac{2J^2}{U} (1+\delta)^2 \sum_{j} \hat{b}_{j}^{\dagger} \hat{b}_{j+1} + \frac{2J^2}{U} (1-\delta)^2 \sum_{j} \hat{b}_{j+1}^{\dagger} \hat{b}_{j}
\end{eqnarray}
\end{widetext}

The effective Hamiltonian after second-order perturbation theory thus takes the form:

\begin{eqnarray} 
\hat{H}_{\mathrm{eff}} &=& U \sum_{j=0}^L \hat{b}_j^{\dagger} \hat{b}_j + \frac{4J^2}{U} (1-\delta^2) \sum_{j=0}^L\hat{b}_{j}^{\dagger} \hat{b}_{j} \nonumber \\
&+& \frac{2J^2}{U} (1-\delta)^2 \sum_{j=0}^{L-1} \hat{b}_{j+1}^{\dagger} \hat{b}_{j} + \frac{2J^2}{U} (1+\delta)^2 \sum_{j=0}^{L-1} \hat{b}_{j}^{\dagger} \hat{b}_{j+1}\nonumber\\
\end{eqnarray}
We note that due to doublon number conservation, the terms proportional to $\sum_j b_j^{\dagger} b_j$ generate a global phase to the wavefunction, and do not affect any physical quantities. Thus up to these additional constant terms ($\propto \sum_j b_j^{\dagger} b_j$), the effective Hamiltonian is given by 
\begin{equation}
{\hat{H}_{\mathrm{eff}} = J_{R} \sum_{j=0}^{L-1} \hat{b}_{j+1}^{\dagger} \hat{b}_{j}  + J_{L} \sum_{j=0}^{L-1} \hat{b}_{j}^{\dagger} \hat{b}_{j+1}},
\end{equation}
where
$J_{R} = \frac{2J^2}{U} (1-\delta)^2$ and $J_{L} = \frac{2J^2}{U} (1+\delta)^2$. We conclude that the doublon QW in the HNBH model can be mapped to an effective single-particle HN QW in the strong interaction regime.
\section{Effective doublon Hamiltonian in
the presence of a field $(F \neq 0)$}
\label{App:B}
In this appendix, we generalize the analysis in App.~\ref{App:A} to the case of a nonzero dc field ($F \ne 0$), where the Hamiltonian is described by:
\begin{align*}
\hat{H} = 
& \sum_{j=0}^{L-1} -J\left[ (1 - \delta)\, \hat{a}_{j+1}^{\dagger} \hat{a}_j + (1 + \delta)\, \hat{a}_j^{\dagger} \hat{a}_{j+1} \right] \nonumber \\
& + \frac{U}{2} \sum_{j=0}^{L} \, \hat{a}_j^{\dagger} \hat{a}_j^{\dagger} \hat{a}_j \hat{a}_j + F \sum_{j=0}^{L} j \, \hat{a}_j^{\dagger} \hat{a}_j,
\end{align*}
We proceed to analyze this system by first transforming the system to a rotating frame through the transformation, $\mathcal{\hat{U}}_{R} = \exp[-iFt \sum_{j} j \hat{n}_{j}]$, such that the rotating-frame Hamiltonian takes the form:
\begin{equation}
    \hat{H}_{\mathrm{rot}}(t) = \mathcal{\hat{U}}_{R}^{\dagger} \hat{H} \mathcal{\hat{U}}_{R} - i\mathcal{\hat{U}}_{R}^{\dagger} \dot{\hat{\mathcal{U}_{R}}}
\label{eq:Hrot}    
\end{equation}
Evaluating Eq.~\ref{eq:Hrot}, we obtain
\begin{widetext}
\begin{equation}
    \hat{H}_{\mathrm{rot}} = - J \sum_{j=0}^{L-1} \left[ (1- \delta)e^{iFt} \hat{a}_{j+1}^{\dagger} \hat{a}_{j} + (1+ \delta) e^{-iFt} \hat{a}_{j}^{\dagger} \hat{a}_{j+1}\right] + \frac{U}{2} \sum_{j=0}^L \hat{a}_{j}^{\dagger}\hat{a}_{j}^{\dagger}\hat{a}_{j}\hat{a}_{j}.
\end{equation}
\end{widetext}
Following the approach outlined in App.~\ref{App:A}, we now expand the effective Hamiltonian in the rotating frame, $\tilde{H}_{\rm eff}$ in the form 
\begin{equation}
\hat{\tilde{H}}_{\rm eff} = \hat{H}^{(0)}_{\rm rot} + \hat{H}^{(1)}_{\rm rot}+ \hat{H}^{(2)}_{\rm rot} + \ldots,
\end{equation}
where 
\begin{equation}
\hat{H}^{(0)}_{\rm rot} = \frac{U}{2} \sum_{j=0}^L \hat{a}_{j}^{\dagger}\hat{a}_{j}^{\dagger}\hat{a}_{j}\hat{a}_{j},  
\end{equation}
remains unchanged under this rotating frame transformation.

Performing the same perturbation theory analysis outlined in App.~\ref{App:A}, we obtain, $H_{\rm rot}^{(1)} = 0$, and 
\begin{widetext}
\begin{equation}
\hat{H}_{\rm rot}^{(2)} = \frac{2J^2}{U} (1-\delta)^2 e^{2iFt}\sum_{j=0}^{L-1}\hat{b}_{j+1}^{\dagger} \hat{b}_{j}
 + \frac{2J^2}{U} (1+\delta)^2 e^{-2iFt} \sum_{j=0}^{L-1}\hat{b}_{j}^{\dagger} \hat{b}_{j+1} + \frac{4J^2}{U} (1-\delta^2) \sum_{j=0}^L\hat{b}_{j}^{\dagger} \hat{b}_{j}
\end{equation}
\end{widetext}
Thus, in the rotating frame, up to additional constant terms ($\propto \sum_j \hat{b}_j^{\dagger} \hat{b}_j$), we obtain a rotating-frame effective Hamiltonian of the form:

\begin{equation}
\hat{\tilde{H}}_{\mathrm{eff}} = J_R^{{\rm (rot)}} \sum_{j=0}^{L-1} \hat{b}_{j+1}^{\dagger} \hat{b}_{j} + J_L^{{\rm (rot)}} \sum_{j=0}^{L-1} \hat{b}_{j}^{\dagger} \hat{b}_{j+1},
\end{equation}
where $J_R^{{\rm (rot)}} = \frac{2J^2}{U} (1-\delta)^2 e^{2iFt}$ and $J_L^{{\rm (rot)}} = \frac{2J^2}{U} (1+\delta)^2 e^{-2iFt}$ respectively.\\

Finally, we note that after transforming back to the laboratory frame, the effective Hamiltonian is given by:

\begin{eqnarray}
\hat{H}_{\rm eff} &=& \frac{2J^2}{U} (1-\delta)^2  \sum_{j=0}^{L-1}  \hat{b}_{j+1}^{\dagger} \hat{b}_{j}  
 + \frac{2J^2}{U} (1+\delta)^2  \sum_{j=0}^{L-1}  \hat{b}_{j}^{\dagger} \hat{b}_{j+1} \nonumber\\
 &+& 2 F \sum_{j=0}^{L}  j \hat{b}_{j}^{\dagger}\hat{b}_{j}.
\end{eqnarray}

We conclude that, in the strong interaction regime, the doublon QW can be mapped to a single-particle HN QW under a dc field, $2F$.

\section{Validity of the Effective Hamiltonians}
\label{App:C}
In Apps.~\ref{App:A} and \ref{App:B}, we have derived the effective Hamiltonians that describe the doublon dynamics in the
$J/U \rightarrow 0$ limit. In this appendix, we discuss their regime of
validity.\\

We begin by noting that the non-reciprocal tunneling can be removed by
the imaginary gauge transformation $\hat{S} = r^{\sum_{j} j\hat{n}_{j}}$
with $r = \sqrt{(1+\delta)/(1-\delta)}$~\cite{hatano1996localization},
under which $\hat{S}\hat{H}\hat{S}^{-1}$ takes the Hermitian BH form
with symmetric tunneling amplitude $J' = J\sqrt{1-\delta^{2}}$. The
interaction and dc-field terms are on-site terms, and are consequently
unaffected. As discussed in Sec.~\ref{sec:Model}, in the Hermitian
limit the two-particle spectrum for $U/J \gg 1$ splits into a lower
band of $L(L-1)/2$ scattering states and an upper band of $L$ doublon
bound states~\cite{lahini2012quantum}. With the renormalized tunneling
amplitude, the scattering band spans $[-4J',\, 4J']$, while the doublon
band is centered at $U + 4J'^{2}/U$ with half-width $4J'^{2}/U$, so
that its lower edge lies at $U$. The two bands are therefore separated
when
\begin{equation}
U > 4J\sqrt{1-\delta^{2}}.
\label{eq:gapcond}
\end{equation}
When Eq.~(\ref{eq:gapcond}) holds, the doublon band is energetically
isolated from the scattering band, so that the doublon-band eigenstates
hybridize only weakly with scattering states and are well described as
dressed doublons, with a dressing controlled by $J/U$. As we show
below, the departure of $P_{\rm doub}$ from $1$ then reflects only
this dressing and the finite overlap of the initial state with the
scattering band.

In order to determine the probability of being in the doublon subspace,
$P_{\rm doub}(t) = \langle \psi(t) \rvert \hat{P}_{G} \lvert \psi(t)
\rangle = \frac{1}{2}\sum_{i}\hat{n}^{(2)}_{i}$ ($\hat{P}_{G}$ is
defined in App.~\ref{App:A}), we first transform to the Hermitian frame
via $\hat{S}$. Since $\hat{H}^{(0)}$ takes the value $U$ on the doublon
subspace and $0$ on its complement, to the first order in $J/U$, each doubly
occupied configuration is dressed as
\begin{equation}
\lvert G_{i} \rangle \;\to\; \lvert G_{i} \rangle
+ \frac{1}{U}\,(1-\hat{P}_{G})\,\hat{V}'\lvert G_{i} \rangle,
\end{equation}
where $\hat{V}'$ is the Hermitian-frame counterpart of the tunneling
term $\hat{V}$. The weight of this admixture is of second order in
$J/U$:
\begin{equation}
\varepsilon = \frac{\langle G_{i} \rvert \hat{P}_{G}\hat{V}'
(1-\hat{P}_{G})\hat{V}'\hat{P}_{G} \lvert G_{i} \rangle}{U^{2}}
= \frac{4J'^{2}}{U^{2}}.
\end{equation}
As we discuss below, this admixture reduces $P_{\rm doub}$ to a value
below $1$.

\begin{figure}
\includegraphics[width=\linewidth]{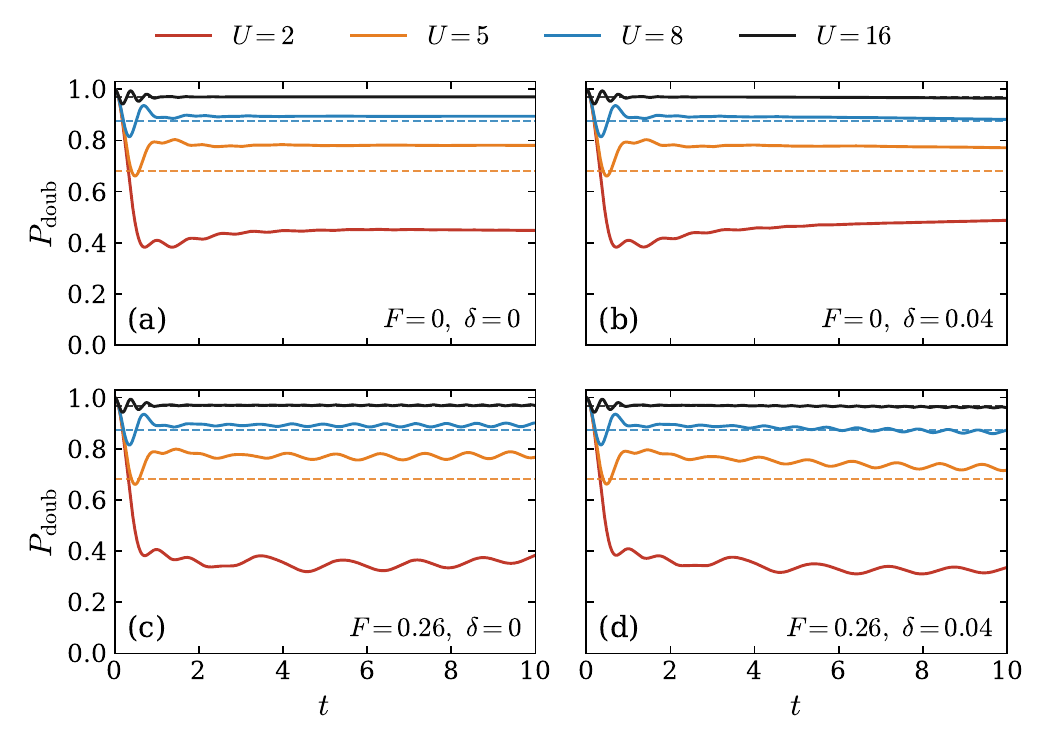}
\caption{The weight in the doublon subspace,
$P_{\rm doub} = \tfrac{1}{2}\sum_{i}\hat{n}^{(2)}_{i}$, for the initial
state $\lvert\psi^{(2)}_{\rm ini}\rangle$ with $L = 70$:
(a) $F = 0$, $\delta = 0$; (b) $F = 0$, $\delta = 0.04$;
(c) $F = 0.26$, $\delta = 0$; (d) $F = 0.26$, $\delta = 0.04$. Dashed
lines show the estimate $1 - 8J^{2}/U^{2}$ [Eq.~(\ref{eq:leakage})] for
$U = 5$, $8$ and $16$. The analytical estimate agrees quantitatively
with the numerical results for $U \gtrsim 8$ and remains qualitatively
accurate at $U = 5$, where higher-order corrections enhance the plateau
value of $P_{\rm doub}$ relative to the estimate. For $U = 2$,
Eq.~(\ref{eq:gapcond}) is violated and the estimate, which becomes
negative, is not shown.}
\label{fig:leakage}
\end{figure}

To see this, we note that the long-time average of $P_{\rm doub}$ is
governed by the diagonal ensemble,
\begin{equation}
\overline{P_{\rm doub}} = \sum_{n} \lvert \langle n \vert
\psi^{(2)}_{\rm ini} \rangle \rvert^{2} \,
\langle n \rvert \hat{P}_{G} \lvert n \rangle,
\end{equation}
where $\{\lvert n \rangle\}$ are the eigenstates in the Hermitian
frame. The admixture enters the long-time average of $P_{\rm doub}$
twice: (a) the initial state, being a bare doublon, carries a weight
$1 - \varepsilon$ on the doublon band and $\varepsilon$ on the
scattering band, and (b) each doublon-band eigenstate is, on average,
doubly occupied only with probability $1 - \varepsilon$, while the
scattering-band eigenstates carry only $O(\varepsilon)$ weight in the
doublon subspace. We therefore obtain, in the Hermitian frame,
\begin{equation}
\overline{P_{\rm doub}} \simeq (1 - \varepsilon)^{2}
\simeq 1 - \frac{8J^{2}(1-\delta^{2})}{U^{2}}.
\end{equation}
The transformation back to the non-Hermitian frame via $\hat{S}^{-1}$ modifies this result at the same order in $\delta$. For the values of $\delta$ considered in this work, these corrections are negligible, and to the leading order we obtain
\begin{equation}
\overline{P_{\rm doub}} \simeq 1 - \frac{8J^{2}}{U^{2}}.
\label{eq:leakage}
\end{equation}

Our analytical prediction agrees well with the exact evolution of
$\lvert\psi^{(2)}_{\rm ini}\rangle$, shown in Fig.~\ref{fig:leakage}. After a short transient, $P_{\rm doub}$ settles to a plateau whose time-averaged value is essentially independent of both $\delta$ and $F$. At $U = 5$ the analytical estimate lies below the numerically obtained doublon weight, indicating that higher-order corrections increase the occupation of the doublon subspace. For $U = 2$, Eq.~(\ref{eq:gapcond}) is violated, and the effective description is no longer expected to hold. Finally, we note that the ratio $J_{R}/J_{L}$, which controls the directionality of the doublon walk, is unaffected at this order in $J/U$.

\bibliography{ref}

\end{document}